\documentstyle[preprint,aps]{revtex}

\begin{document}
\title{Dynamical Exchange Effects in a Two-Dimensional Many-Polaron Gas}
\author{K. J. Hameeuw, J. Tempere, F. Brosens, J. T. Devreese}
\address{TFVS, Universiteit Antwerpen, Universiteitsplein 1, B2610
Antwerpen, Belgium.}
\date{\today }
\maketitle

\begin{abstract}
We calculate the influence of dynamical exchange effects on the response
properties and the static properties of a two-dimensional many-polaron gas.
These effects are not manifested in the random-phase approximation which is
widely used in the analysis of the many-polaron system. Here they are taken
into account by using a dielectric function derived in the time-dependent
Hartree-Fock formalism. At weak electron-phonon coupling, we find that
dynamical exchange effects lead to substantial corrections to the
random-phase approximation results for the ground state energy, the
effective mass, and the optical conductivity of the polaron system.
Furthermore, we show that the reduction of the spectral weight of the
optical absorption spectrum at frequencies above the longitudinal optical
phonon frequency, due to many-body effects, is overestimated by the
random-phase approximation.
\end{abstract}


\section{Introduction}

\tighten Although the static and dynamic properties of a single polaron have
been \cite{PolaronOld,DevreesePRB5} (and still are \cite{PolaronNew})
extensively studied, the properties of an interacting polaron system, e.g.
their optical absorption \cite{Cataudellaetal,DevreeseinAvellaetal}, are
less well understood. Thus far, the many-polaron system has been studied
mainly in the Hartree-Fock approximation and in the random-phase
approximation (RPA) \cite{LemmensPSS82,WuPRB34,DeganiPRB35,JalabertPRB40},
even though in Ref. \cite{daCostaPRB47} it was shown that in three
dimensions (3D) the polaron energy and effective mass are influenced by
screening effects beyond RPA. In this paper, we investigate dynamical
exchange effects in both static properties (energy and effective mass) and
response properties (optical absorption) of the two-dimensional (2D) polaron
system.

However, although our formalism is appropriate for the description of an
interacting electron (or polaron) gas, it does not allow to investigate the
2D Wigner solid phase formed for example by electrons on a helium surface.
Also the study of the many small-polaron gas and bipolarons in quasi-2D
systems such as the copper oxide planes in high-T$_{c}$ cuprate
superconductors \cite%
{EwardsMottAlexandrov,AlexandrovPRcuprates,AlexandrovTcbipolarons,Alexandrovbook}
lies beyond the scope of our approach in its current form.

A convenient formalism to take into account many-body effects in the polaron
system relies on containing these effects in the structure factor \cite%
{LemmensPSS82}. This approach, reminiscent of the Feynman-Bijl treatment of
superfluid helium-4, was first developed to calculate the ground state
energy of a many-polaron system \cite{LemmensPSS82}, and has later been
applied to the optical absorption of this system \cite{TemperePRB64}. The 2D
dynamic structure factor can be written as 
\begin{equation}
S_{\text{2D}}\left( q,\omega \right) =-\frac{\hbar }{nv_{q}}%
\mathop{\rm Im}%
\left( \frac{1}{\varepsilon \left( q,\omega \right) }\right) ,
\label{strucfac}
\end{equation}%
with $v_{q}=e^{2}/(2\varepsilon _{\infty }q)$ the Fourier transform of the
bare Coulomb interaction in 2D, $n$ the surface density of charge carriers, $%
\varepsilon _{\infty }$ the dielectric constant at high frequency, and $%
\varepsilon (q,\omega )$ the dielectric function of the system of charge
carriers. Previous work \cite%
{LemmensPSS82,WuPRB34,DeganiPRB35,JalabertPRB40,TemperePRB64} relied on the
RPA dielectric function: 
\begin{equation}
\varepsilon _{\text{RPA}}(q,\omega )=1+Q_{0}(q,\omega ),
\end{equation}%
where $Q_{0}(q,\omega )$ is the 2D Lindhard polarizability and is known in
closed form \cite{SternPRL18}. A more accurate result for the dielectric
function is obtained by introducing the frequency dependent local field
correction $G(q,\omega ):$%
\begin{equation}
\varepsilon (q,\omega )=1+\frac{Q_{0}(q,\omega )}{1-G(q,\omega
)Q_{0}(q,\omega )}.
\end{equation}%
In the 3D case, the local field correction was derived in the framework of
the dynamical exchange decoupling (DED) method \cite{BrosensPSS74}. This
method, based on the time-dependent Hartree-Fock formalism, was recently
extended to the 2D case in Refs. \cite{HameeuwSSC126,HameeuwEPJB35}. This
leads to an improved dielectric function, $\varepsilon _{\text{DED}%
}(q,\omega )$, that can be used to include dynamical exchange effects in the
analysis of the many-polaron system. The DED result for the dielectric
function is markedly different from the results obtained in the RPA
formalism, Hubbard's formalism \cite{HubbardPRSL243} and the formalism of
Singwi {\it et al.} \cite{SingwiPR176}. In 3D these different approaches can
be compared to experiment: measurements of the dynamic local field
correction by Larson and co-workers \cite{LarsonPRL77} have shown that the
3D DED approach \cite{BrosensPSS74} leads to superior results.

In this work, we use the DED dielectric function in combination with the
formalism describing the many-body effects in a polaron system through the
structure factor. In section II, we investigate the dynamical exchange
effects on the optical absorption of the many-polaron system. From the
optical conductivity the effective mass of the polaron as a function of the
density is derived through a sum-rule in section III. Finally, in section
IV, the ground state energy of the interacting polaron system is calculated
beyond\ RPA. As we shall show, including dynamical exchange effects leads to
non-negligible corrections to all these quantities.

\section{Optical absorption of the interacting 2D polaron system}

The optical absorption of an interacting system of Fr\"{o}hlich polarons in
the weak-coupling regime was investigated in \cite{WuPRB34,TemperePRB64}.
The optical conductivity in 2D is given by \cite{TemperePRB64} 
\begin{equation}
\mathop{\rm Re}%
[\sigma _{\text{2D}}(\omega )]=\alpha \frac{ne^{2}}{4\omega ^{3}}%
\displaystyle\int %
\limits_{0}^{\infty }dq\text{ }q^{2}S_{\text{2D}}(q,\omega -\omega _{\text{LO%
}}),  \label{Opticon}
\end{equation}%
where $e$ is the electron charge, $\omega _{\text{LO}}$ is the longitudinal
optical (LO) phonon frequency and $\alpha $ is the dimensionless Fr\"{o}%
hlich coupling constant determining the strength of the electron-phonon
interaction. Note that a delta-function peak at $\omega =0,$ omitted in (\ref%
{Opticon}), is present in the optical absorption. The influence of the Fermi
statistics and the screened Coulomb interactions between the electrons are
taken into account through the structure factor $S_{\text{2D}}$ appearing in
expression (\ref{Opticon}). Thus far, the effect of electron-electron
interactions on the {\it optical absorption }of a many-polaron gas has been
investigated only in the framework of RPA \cite%
{WuPRB34,TemperePRB64,IadonisiEPJB12}. Nevertheless, expression (\ref%
{Opticon}) is not restricted to the RPA approximation: any form of the
dynamic structure factor can be introduced in the integrand. Here, we will
use the results obtained for the dielectric function in the DED formalism %
\cite{HameeuwEPJB35} instead.

Using expression (\ref{strucfac}) and the DED result for the dielectric
function, we can rewrite (\ref{Opticon}) as%
\begin{equation}
\mathop{\rm Re}%
\left[ \sigma _{\text{2D}}\left( \nu \right) \right] =-\alpha 
{\displaystyle{{\cal N} \over (\nu /\nu _{LO})^{3}}}%
\displaystyle\int %
\limits_{0}^{\infty }dk\text{ }k^{3}%
\mathop{\rm Im}%
\left( 
{\displaystyle{1 \over \varepsilon _{\text{DED}}\left( k,\nu -\nu _{\text{LO}}\right) }}%
\right) ,  \label{Opticon2}
\end{equation}%
where now $k=q/k_{F}$ and $v=\hbar \omega /(2E_{F})$ are dimensionless wave
numbers and frequencies based on the Fermi wave vector $k_{F}$ and the Fermi
energy $E_{F}$, respectively. The prefactor 
\begin{equation}
{\cal N}=%
{\displaystyle{n\left( \hbar \omega _{LO}\right) ^{2}e^{2} \over 2m_{b}^{2}\omega _{\text{LO}}^{3}}}%
\sqrt{%
{\displaystyle{\hbar  \over 2m_{b}\omega _{\text{LO}}}}%
}%
{\displaystyle{4\pi \varepsilon _{\infty }k_{F}^{2} \over e^{2}}}%
\end{equation}%
contains material constants (the band mass $m_{b}$, the LO phonon frequency $%
\omega _{LO}$ and the high-frequency permittivity $\varepsilon _{\infty }$)
and has dimensions of optical conductivity. Expression (\ref{Opticon2}) for
the optical absorption allows to easily substitute available results for the
dielectric function.

The DED structure factor can be interpreted as a sum of two contributions,
namely a contribution from the single-particle excitations (the Landau
continuum) and a contribution from the plasmon excitations. We can write 
\begin{equation}
\mathop{\rm Im}%
\left[ 
{\displaystyle{1 \over \varepsilon _{\text{DED}}\left( k,\nu \right) }}%
\right] =-A_{\text{pl}}(k)\delta \lbrack \nu -\nu _{\text{pl}}(k)]+%
\mathop{\rm Im}%
\left[ 
{\displaystyle{1 \over \varepsilon _{\text{cont}}\left( k,\nu \right) }}%
\right] ,  \label{splitup}
\end{equation}%
where $\nu _{\text{LO}}$ is the LO phonon frequency in units of $%
2E_{F}/\hbar $ and $\nu _{\text{pl}}(k)$ is the plasmon dispersion defined
by $\varepsilon (k,\nu _{\text{pl}})=0$. The plasmon excitations in our
approach lead to delta-functions, with a strength $A_{\text{pl}}(k)$, in the
frequency dependence of the structure factor. Also the optical conductivity
can be separated in a contribution due to the single-particle excitations
and a part coming from the plasmon branch: 
\begin{equation}
\mathop{\rm Re}%
[\sigma _{\text{2D}}(\nu )]=%
\mathop{\rm Re}%
[\sigma _{\text{2D}}^{\text{pl}}(\nu )]+%
\mathop{\rm Re}%
[\sigma _{\text{2D}}^{\text{cont}}(\nu )],
\end{equation}%
with 
\begin{eqnarray}
\mathop{\rm Re}%
[\sigma _{\text{2D}}^{\text{pl}}(\nu )] &=&\alpha 
{\displaystyle{{\cal N} \over (\nu /\nu _{\text{LO}})^{3}}}%
\frac{k_{0}^{3}A_{\text{pl}}(k_{0})}{\nu _{\text{pl}}^{\prime }(k_{0})}, \\
\mathop{\rm Re}%
[\sigma _{\text{2D}}^{\text{cont}}(\nu )] &=&-\alpha 
{\displaystyle{{\cal N} \over (\nu /\nu _{\text{LO}})^{3}}}%
\displaystyle\int %
\limits_{0}^{\infty }dk\text{ }k^{3}%
\mathop{\rm Im}%
\left( 
{\displaystyle{1 \over \varepsilon _{\text{cont}}\left( k,\nu -\nu _{LO}\right) }}%
\right) .
\end{eqnarray}%
Here, $k_{0}$ is the wave vector at which $\nu -\nu _{\text{LO}}=\nu _{\text{%
pl}}(k)$, and $\nu _{\text{pl}}^{\prime }=d\nu _{\text{pl}}/dk$ is the
derivative of the plasmon frequency with respect to the wave vector. In the
2D electron gas, the plasmon branch $\nu _{\text{pl}}(k)$ lies close to the
edge of the Landau continuum, $k^{2}/2+k$. Figure 1 compares the plasmon
dispersion in the DED approach to the dispersion in the RPA approach for
several values of $r_{s}$. For small $k$ values the plasmon frequency in DED
is lower than in RPA. Furthermore, in DED longer wave length plasmons are
present. The oscillator strength $A_{\text{pl}}(k)$ of the plasmon branch
can be obtained straightforwardly using the $f$-sum rule. If we substitute (%
\ref{splitup}) in the $f$-sum rule, 
\begin{equation}
-%
\displaystyle\int %
\limits_{0}^{\infty }\nu 
\mathop{\rm Im}%
\left( \frac{1}{\varepsilon _{\text{DED}}\left( k,\nu \right) }\right) d\nu =%
{\displaystyle{\pi  \over 2}}%
{\displaystyle{r_{s}k \over \sqrt{2}}}%
,
\end{equation}%
we can find an expression for the oscillator strength of the plasmon branch: 
\begin{equation}
A_{\text{pl}}(k)=%
{\displaystyle{1 \over \nu _{\text{pl}}(k)}}%
\left( 
{\displaystyle{\pi  \over 2}}%
{\displaystyle{r_{s}k \over \sqrt{2}}}%
-M_{1}\right) ,
\end{equation}%
where 
\begin{equation}
M_{1}=-%
\displaystyle\int %
\limits_{0}^{\infty }\nu 
\mathop{\rm Im}%
\left( \frac{1}{\varepsilon _{\text{cont}}\left( k,\nu \right) }\right) d\nu
,
\end{equation}%
is the first frequency moment of the Landau continuum contribution only. In
Figure 2, we compare the oscillator strength of the plasmon branch in RPA
and DED for several values of $r_{s}$. At small $k$ the difference between
the strength of the plasmons in RPA and DED is negligible. As $k$ increases,
a difference appears: the maximum oscillator strength for the plasmon branch
in DED is reached at a smaller wave vector. Compared to RPA, the oscillator
strength of the DED plasmons shows a long tail as a function of $k$.

\bigskip

In Figure 3 the optical absorption due to the plasmon contribution is shown
for several relevant values of the density $n$ of a two dimensional polaron
gas, using the material parameters for GaAs ($\varepsilon _{\infty }=10.8$, $%
m_{\text{b}}=0.0657$ $m_{\text{e}}$, $\hbar \omega _{\text{LO}}=36.77$ meV, $%
a_{\text{Bohr}}=8.78$ nm). The presence of the shorter wave length plasmons
in DED as compared to RPA shifts the spectral weight to higher frequencies.
Nevertheless, both in RPA and DED the contribution of plasmons to the
optical absorption of the many-polaron gas is small in comparison with the
contribution of the single-particle excitations. The total optical
absorption is shown in Figure 4, again for different surface densities of
polarons in GaAs. Although the Landau continuum occupies the same region of
the $\{k,\nu \}$-plane in the DED and RPA approaches, the spectral weight of
the structure factor is distributed differently. As can be seen in Figure 4,
this leads to a measurable difference between the DED and RPA results.
Comparing both the DED and RPA optical absorption with the single-polaron
absorption \cite{note1}, we find that the RPA result overestimates the
damping due to many-body effects.

\bigskip

\section{Effective mass of the polaron in DED}

From Figure 4, it is clear that the spectral weight of the optical
absorption in the region beyond the single-phonon gap $(\omega >\omega _{%
\text{LO}})$ is different for each of the approaches (DED, RPA and
single-polaron). This may seem surprising at first glance, since the $f$-sum
rule imposes that the total spectral weight is a constant: 
\begin{equation}
\displaystyle\int %
\limits_{0}^{\infty }%
\mathop{\rm Re}%
[\sigma (\omega )]d\omega =\frac{\pi Ne^{2}}{2m_{b}}.
\end{equation}%
It is important to keep in mind that the optical absorption of the polaron
system shows a delta-function in the origin ($\omega =0$), not shown in
Figure 4. The total spectral weight is distributed between this
delta-function peak and the optical absorption spectrum beyond the
single-phonon gap \cite{DevreesePRB15}. The spectral weight of the
delta-function is related to the polaron effective mass $m^{\ast }$ in such
a way that \cite{DevreesePRB15} 
\begin{equation}
\displaystyle\int %
\limits_{\omega >\omega _{LO}}^{\infty }%
\mathop{\rm Re}%
[\sigma (\omega )]d\omega =\frac{\pi Ne^{2}}{2m_{b}}\left( 1-\frac{m_{b}}{%
m^{\ast }}\right) .  \label{sumrule}
\end{equation}%
Thus, a change in the spectral weight allotted to the $\omega >\omega _{LO}$
region of the optical conductivity indicates a change in the effective mass
of the polarons. The electron-phonon coupling increases the effective mass
of a single polaron by $\Delta m=m^{\ast }-m_{b}$. The well-known
single-polaron result for this effective mass increase in 2D is $\Delta
m/m_{b}=\alpha \pi /8$. In Figure 5, we show the effective mass increase $%
\Delta m/(\alpha m_{b})=(m^{\ast }/m_{b}-1)/\alpha $ as a function of the
surface density of polarons in GaAs. Both the RPA result (dashed curve, cf. %
\cite{WuPRB34,DeganiPRB35,JalabertPRB40,TempereEPJB20}) and the DED result
(full curve) approach the one-polaron result in the limit of small density.
As the surface density of polarons increases, $\Delta m$ decreases, more
rapidly in RPA than in DED.

The fractional change $\Delta m_{\text{DED}}/\Delta m_{\text{RPA}}$ is shown
in the inset of Figure 5, and indicates that there is roughly a 15\%
difference between the effective mass increase predicted by DED and that
predicted by RPA. This is consistent with the increase in spectral weight of
the optical absorption at $\omega >\omega _{\text{LO}}$ in DED as compared
to RPA: if less spectral weight is present in the region $\omega >\omega _{%
\text{LO}}$, the delta-function in the origin must carry more spectral
weight, indicating an increase in effective mass.

\bigskip

\section{Ground-state energy of the interacting 2D polaron system}

A general expression for the ground state energy per particle of an
interacting polaron gas at weak coupling was derived in Ref. \cite%
{LemmensPSS82} :%
\begin{equation}
E=E_{\text{el}}-\sum_{{\bf q}}\frac{|V_{{\bf q}}|^{2}S_{\text{2D}}({\bf q})}{%
\hbar \omega _{LO}+%
{\displaystyle{(\hbar q)^{2} \over 2m_{b}S_{\text{2D}}({\bf q})}}%
}.
\end{equation}%
In this expression $E_{\text{el}}$ is the energy contribution per particle
of the system of charge carriers, without electron-phonon coupling, and the
second term is the polaron contribution. The 2D Fr\"{o}hlich interaction
amplitude appearing in this expression is%
\begin{equation}
\left| V_{{\bf q}}\right| ^{2}=\frac{(\hbar \omega _{LO})^{2}}{q}\frac{2\pi
\alpha }{A}\sqrt{\frac{\hbar }{2m_{b}\omega _{LO}}},
\end{equation}%
with $A$ the surface of the 2D system. Whereas the optical conductivity
depends on the dynamic structure factor, the ground state energy depends on
the static structure factor%
\begin{eqnarray}
S_{\text{2D}}(q) &=&%
\displaystyle\int %
\limits_{0}^{\infty }\frac{d\omega }{\pi }S_{\text{2D}}(q,\omega )  \nonumber
\\
&=&-\frac{\hbar }{n\pi v_{q}}%
\displaystyle\int %
\limits_{0}^{\infty }%
\mathop{\rm Im}%
\left( \frac{1}{\varepsilon \left( q,\omega \right) }\right) d\omega .
\end{eqnarray}%
Expressed in dimensionless variables for the wave number and the frequency,
this is%
\begin{equation}
S_{\text{2D}}(k)=-\frac{\sqrt{2}k}{\pi r_{s}}%
\displaystyle\int %
\limits_{0}^{\infty }%
\mathop{\rm Im}%
\left( \frac{1}{\varepsilon \left( k,\nu \right) }\right) d\nu .
\end{equation}%
A comparison between the static structure factor in RPA and DED is shown in
Figure 6. Around $k=2,$ the DED static structure factor shows a transition
to the long wave length behavior \cite{HameeuwSSC126,HameeuwEPJB35}.

The shift in ground state energy per particle due to the electron-phonon
coupling $\Delta E_{\text{pol}}=E(\alpha )-E(\alpha =0)$ is given by \cite%
{LemmensPSS82}: 
\begin{equation}
\Delta E_{\text{pol}}=-\alpha \hbar \omega _{LO}%
\displaystyle\int %
\limits_{0}^{\infty }dk\text{ }\gamma \frac{\lbrack S_{\text{2D}}(k)]^{2}}{%
S_{\text{2D}}(k)+\gamma ^{2}k^{2}}.
\end{equation}%
with $\gamma =k_{F}/k_{LO}$ and $k_{LO}=\sqrt{2m_{b}\omega _{LO}/\hbar }$.
The result for the polaronic shift in the ground state energy per particle,
as a function of the Wigner-Seitz radius $r_{s}$, is shown in Figure 7 for
different approaches. The material parameters for GaAs were used for this
figure. If no many-body effects are taken into account, the 2D polaronic
energy per particle is $-(\pi /2)\alpha \hbar \omega _{LO}$ in the weak
coupling regime. If many-body effects are taken into account on the level of
the Hartree-Fock approximation, we get a monotonous decreasing energy as $%
r_{s}$ increases, while the RPA approximation leads to a minimum in the
polaronic energy at $r_{s}=1.59$, with $E/N=-0.70$ $\alpha \hbar \omega
_{LO} $. We find that including dynamical exchange effects lowers $\left|
\Delta E_{\text{pol}}\right| $ with respect to RPA, and shifts the energy
minimum to slightly lower density: $r_{s}=1.61$, where $E/N=-0.77$ $\alpha
\hbar \omega _{LO}$. So, using an improved dielectric function as compared
to RPA leads to a change of ca. 10 \% in the ground state energy.

\bigskip

\section{Conclusions}

Efforts to describe many-body effects in the polaron system have thus far
relied strongly on the RPA approach \cite%
{LemmensPSS82,WuPRB34,DeganiPRB35,JalabertPRB40,IadonisiEPJB12,TemperePRB64}%
, even though there have been indications that inclusion of a {\sl static}
local field correction modifies the ground state energy and the effective
mass of the polarons \cite{daCostaPRB47}. In this paper, we use a dielectric
function that includes a {\sl frequency-dependent} local field correction
derived in the time-dependent Hartree-Fock formalism \cite%
{BrosensPSS74,HameeuwSSC126,HameeuwEPJB35}. We find that taking into account
dynamical exchange effects reveals important corrections at weak coupling to
the optical absorption of the polaron system, the effective mass of the
polarons and the ground state energy of the polaron system. In GaAs,
estimates of the effective mass based on RPA are roughly 15\% off from the
improved results, estimates of the ground state energy differ roughly 10\%.
Furthermore, we show that in the optical absorption, RPA considerably
overestimates the reduction of spectral weight due to the many-body effects.

\bigskip

\section{Acknowledgements}

The authors thank D. Saeys for pointing out an error in the numerical
calculations of the plasmon contribution to the optical absorption.
Discussions with P. Calvani are gratefully acknowledged. Two of the authors
(K. J. H. and J. T.) are supported financially by the Fund for Scientific
Research - Flanders (Fonds voor Wetenschappelijk Onderzoek -- Vlaanderen).
This research has been supported financially by the FWO-V projects Nos.
G.0435.03, G.0306.00, the W.O.G. project WO.025.99N.and the GOA BOF UA 2000,
IUAP.

\bigskip

\bigskip

\begin{figure}[tbp]
\caption{The plasmon dispersion $\protect\nu _{\text{pl}}(k)$ in a 2D
electron gas is shown as a function of the wave number $k$ (in units of $%
k_{F}$), relative to the edge of the Landau continuum $k^{2}/2+k$, for
different values of $r_{s}$. The full curves correspond to the DED result,
and the dashed curves to the RPA result. In the inset, the location of the
plasmon branch and the Landau continuum (hatched area) are shown.}
\end{figure}

\bigskip

\begin{figure}[tbp]
\caption{The oscillator strength $A_{\text{pl}}\left( k\right) $ of the
plasmon excitation in a 2D electron gas at $\protect\nu _{\text{pl}}(k)$ is
shown as a function of the wave number $k$ (in units of $k_{F}$), for
different values of $r_{s}$. The full curves correspond to the DED result,
the dashed curves to the RPA result.}
\end{figure}

\bigskip

\begin{figure}[tbp]
\caption{The contribution of the plasmon excitations to the optical
absorption of the many-polaron gas is shown as a function of the frequency $%
\protect\nu $ in units of the LO phonon frequency $\protect\nu _{\text{LO}}$%
, for $r_{s}=0.64$ (corresponding to $n=10^{12}$ cm$^{-2}$), $r_{s}=0.91$ ($%
n=5\times 10^{11}$ cm$^{-2}$), and $r_{s}=1.17$ ($n=3\times 10^{11}$ cm$%
^{-2} $). Full curves represent the DED result, and dashed curves the RPA
result.}
\end{figure}

\bigskip

\begin{figure}[tbp]
\caption{The total optical absorption of a 2D many-polaron system in GaAs is
shown as a function of the frequency $\protect\nu $ in units of the LO
phonon frequency $\protect\nu _{\text{LO}}$, for $r_{s}=0.64$ (corresponding
to $n=10^{12}$ cm$^{-2}$) and $r_{s}=1.17$ ($n=3\times 10^{11}$ cm$^{-2}$).
Results are shown for the DED formalism (crosses), the RPA formalism (dashed
curves) and for the single-polaron case\ (dotted curves). The noise present
in the results is due to the numerical treatment. The delta-function in the
optical absorption at $\protect\nu =0$ is not shown in this figure.}
\end{figure}

\bigskip

\begin{figure}[tbp]
\caption{The difference between the polaronic effective mass and the band
mass $\Delta m=m^{\ast }-m_{b}$ is shown as function of the surface density $%
n$ of the 2D polaron system in GaAs. The $r_{s}$ values corresponding to the
densities $n$ are reported on the top axis. The DED result is plotted as a
full curve, the RPA result as a dashed curve, and the one-polaron result as
a dotted line. The inset shows the ratio between the effective mass
enhancement in DED and RPA, as a function of density.}
\end{figure}

\bigskip

\begin{figure}[tbp]
\caption{The static structure factor of a 2D electron gas is shown as a
function of the wave number $k$ (in units of $k_{F}$), in the Hartree-Fock
approximation (dotted curve), in RPA (dashed curves, at $r_{s}=0.5$ and $2.0$%
), and in DED (full curves, at $r_{s}=0.5$ and $2.0$).}
\end{figure}

\bigskip

\begin{figure}[tbp]
\caption{The polaronic ground state energy in a 2D polaron system in GaAs is
plotted as a function of $r_{s}$ in the Hartree-Fock approximation
(dash-dotted curve), in RPA (dashed curve) and in DED (full curve). The
dotted line indicates the one-polaron result.}
\end{figure}

\bigskip


\bigskip

\begin{references}
\bibitem{PolaronOld} For a review, see e.g. J. T. Devreese, in {\it %
Encyclopedia of Applied Physics}, edited by G. L. Trigg (VCH, New York,
1996), Vol. {\bf 14}, 6440 (1995) and references therein.

\bibitem{DevreesePRB5} J.T. Devreese, J. De Sitter, M. Goovaerts, Phys. Rev.
B {\bf 5}, 2367 (1972).

\bibitem{PolaronNew} A. S. Mishchenko, N. Nagaosa, N.V. Prokof'ev, A.
Sakamoto, and B.V. Svistunov, Phys.\ Rev. Lett. {\bf 91}, 236401 (2003).

\bibitem{Cataudellaetal} V. Cataudella, G. De Filippis, and G. Iadonisi,
Eur. Phys. J. B {\bf 12}, 17 (1999).

\bibitem{DevreeseinAvellaetal} J.T. Devreese, {\it Polarons}, in: {\it %
Lectures on the Physics of Highly Correlated Electron Systems VII}, pp.
3-56, Eds. A.Avella and F. Mancini, AIP, New-York (2003).

\bibitem{LemmensPSS82} L. F. Lemmens, J. T. Devreese, and F. Brosens, Phys.
Stat. Sol. (b) {\bf 82}, 439 (1977).

\bibitem{WuPRB34} X. Wu, F. Peeters, and J. T. Devreese, Phys. Rev. B {\bf 34%
}, 2621 (1986).

\bibitem{DeganiPRB35} M. H. Degani and O. Hip\'{o}lito, Phys. Rev. B {\bf 35}%
, 7717 (1987).

\bibitem{JalabertPRB40} R. Jalabert and S. Das Sarma, Phys. Rev. B {\bf 39},
5542 (1989); R. Jalabert and S. Das Sarma, Phys. Rev. B {\bf 40}, 9723
(1989).

\bibitem{daCostaPRB47} W. B. da Costa and N. Studart, Phys. Rev. B {\bf 47},
6356 (1993).

\bibitem{EwardsMottAlexandrov} P.P. Edwards, N.F. Mott, and A.S. Alexandrov,
Journal of Superconductivity {\bf 11}, 151 (1998).

\bibitem{AlexandrovPRcuprates} A. S. Alexandrov, Phys. Rev. B {\bf 61},
12315 (2000).

\bibitem{AlexandrovTcbipolarons} A.S. Alexandrov, Int. J. Mod. Phys. B {\bf %
17}, 3315 (2003).

\bibitem{Alexandrovbook} A. S. Alexandrov and N. F. Mott, Rep. Prog. Phys. 
{\bf 57}, 1197 (1994); see also {\it Theory of Superconductivity: from Weak
to Strong Coupling}, A. S. Alexandrov (IOP publishing, Bristol, UK, 2003).

\bibitem{TemperePRB64} J.\ Tempere, J. T. Devreese, Phys. Rev. B {\bf 64},
104504 (2001).

\bibitem{SternPRL18} F. Stern, Phys. Rev. Lett. {\bf 18}, 546 (1967).

\bibitem{BrosensPSS74} F. Brosens, L.F. Lemmens, J.T. Devreese, Phys. Stat.
Sol. (b) {\bf 74}, 45 (1976); J.T. Devreese, F. Brosens, L.F. Lemmens, Phys.
Rev. B {\bf 21}, 1349 (1980); F. Brosens, J.T. Devreese, L.F. Lemmens,
Phys.\ Rev. B {\bf 21}, 1363 (1980).

\bibitem{HameeuwSSC126} K. J. Hameeuw, F. Brosens, and J. T. Devreese, Solid
State Commun. {\bf 126}, 695 (2003).

\bibitem{HameeuwEPJB35} K. J. Hameeuw, F. Brosens, and J. T. Devreese, Eur.
Phys. J. B {\bf 35}, 93 (2003).

\bibitem{HubbardPRSL243} J. Hubbard, Proc. Roy. Soc. London Ser. A {\bf 243}%
, 336 (1957).

\bibitem{SingwiPR176} K. S. Singwi, M. P. Tosi, R. H. Land, and A. Sj\"{o}%
lander, Phys. Rev. {\bf 176}, 589 (1968).

\bibitem{LarsonPRL77} B.C. Larson, J.Z. Tischler, E.D. Isaacs, P. Zschack,
A. Fleszar, A.G. Eguiluz, Phys. Rev. Lett. {\bf 77}, 1346 (1996); J.Z.
Tischler, B.C. Larson, P. Zschack, A. Fleszar A.G. Eguiluz, Phys. Stat. Sol.
(b) {\bf 237}, 280 (2003).

\bibitem{IadonisiEPJB12} V. Cataudella, G. De Filippis, and G. Iadonisi,
Eur. Phys. J. B {\bf 12}, 17 (1999).

\bibitem{note1} Note that the single-polaron results can be obtained by
using an effective loss function $-%
\mathop{\rm Im}%
[1/\varepsilon ]=[\pi r_{s}/(\sqrt{2}k)]\delta (\nu -k^{2}/2)$. The
delta-function fixes the free-particle dispersion, whereas the
proportionality factor ensures that the sum rules are satisfied.

\bibitem{DevreesePRB15} J.T. Devreese, L.F. Lemmens, and J. Van Royen, Phys.
Rev. B {\bf 15}, 1212 (1977).

\bibitem{TempereEPJB20} J. Tempere and J. T. Devreese, Eur. Phys. J. B {\bf %
20}, 27 (2001).
\end{references}
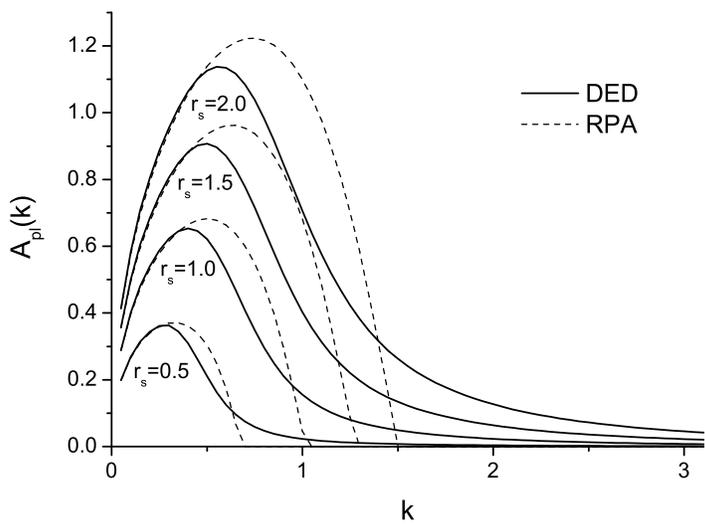
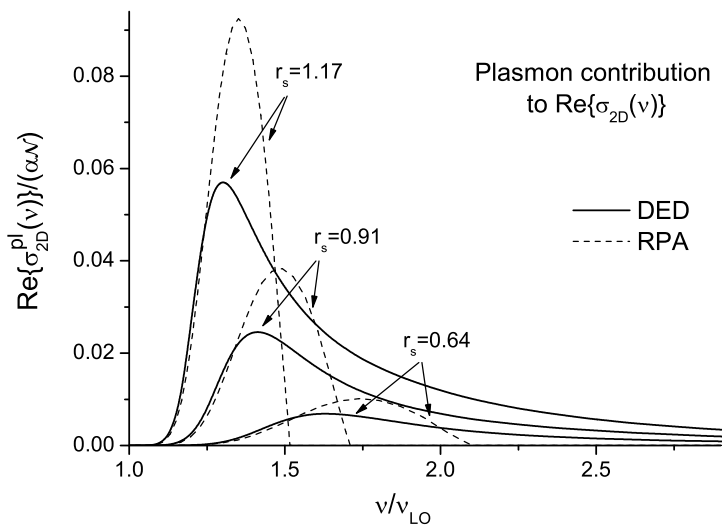
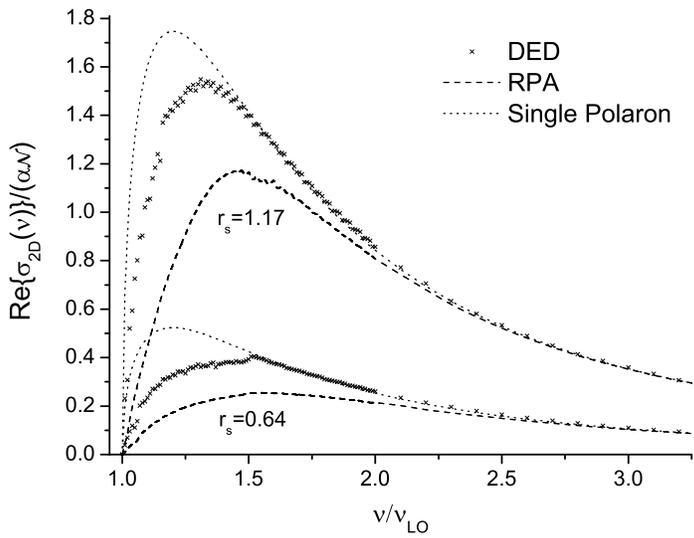
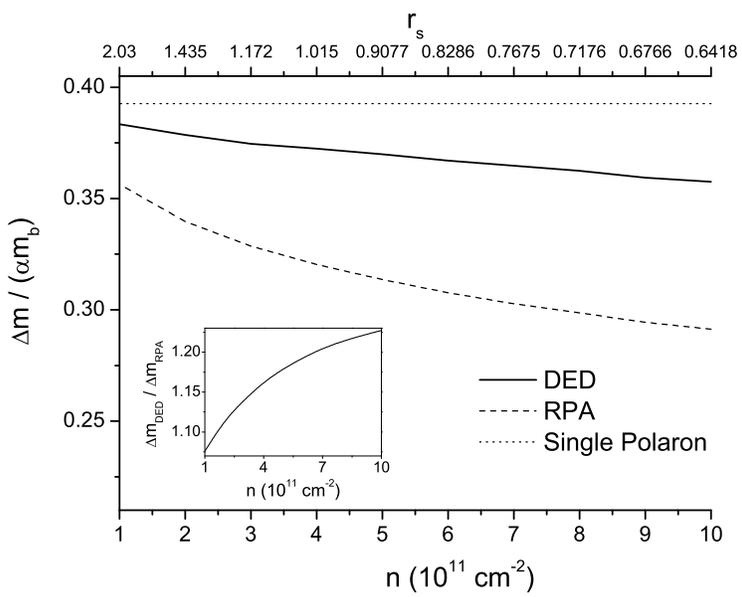
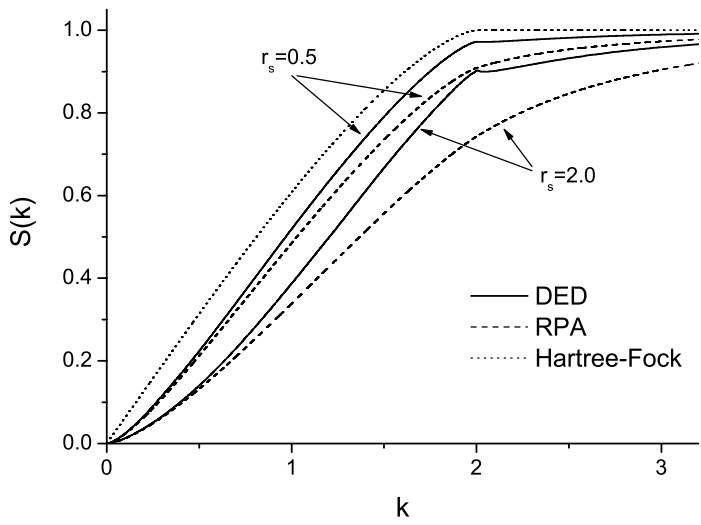
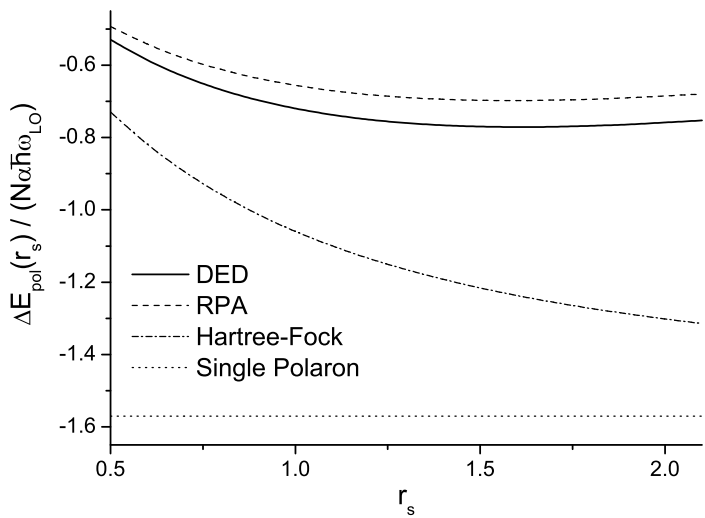
\end{document}